\documentclass{ws-mpla}
\usepackage[super,compress]{cite}
\usepackage{hyperref}
\hypersetup{
  colorlinks=true,        
  linkcolor=blue,         
  citecolor=red,         
}

\usepackage{graphicx}
\begin{document}

%
\catchline{}{}{}{}{}
%

\title{Comments on ``Casimir Effect in the Kerr spacetime with Quintessence"}

\author{BOBIR TOSHMATOV}

\address{Institute of Physics and Research Centre of Theoretical Physics and Astrophysics,\\ Faculty of Philosophy \& Science, Silesian University in Opava, Bezru\v{c}ovo n\'{a}m\v{e}st\'{i} 13,  CZ-74601 Opava, Czech Republic\\
Institute of Nuclear Physics, Ulughbek, Tashkent 100214, Uzbekistan\\
bobir.toshmatov@fpf.slu.cz}

\author{ZDEN\v{E}K STUCHL\'{I}K }

\address{Institute of Physics and Research Centre of Theoretical Physics and Astrophysics,\\ Faculty of Philosophy \& Science, Silesian University in Opava, Bezru\v{c}ovo n\'{a}m\v{e}st\'{i} 13,  CZ-74601 Opava, Czech Republic\\
zdenek.stuchlik@fpf.slu.cz}

\author{BOBOMURAT AHMEDOV}

\address{Ulugh Beg Astronomical Institute, Astronomicheskaya 33, Tashkent 100052, Uzbekistan\\
Institute of Nuclear Physics, Ulughbek, Tashkent 100214, Uzbekistan\\
National University of Uzbekistan, Tashkent 100174, Uzbekistan\\
ahmedov@astrin.uz}

\maketitle

\begin{history}
\received{Day Month Year}
\revised{Day Month Year}
\end{history}

\begin{abstract}
This comment is devoted to the recalculation of the Casimir energy of a massless scalar field in the Kerr black hole surrounded by quintessence derived in [B. Toshmatov, Z. Stuchl\'{i}k and B. Ahmedov, Eur. Phys. J. Plus {\bf 132}, 98 (2017)] and its comparison with the results recently obtained in [V. B. Bezerra, M. S. Cunha, L. F. F. Freitas and C. R. Muniz, Mod. Phys. Lett. A {\bf 32}, 1750005 (2017)] in the spacetime [S. G. Ghosh, Eur. Phys. J. C {\bf 76}, 222 (2016)]. We have shown that in the more realistic spacetime which does not have the failures illustrated here, the Casimir energy is significantly bigger than that derived in [V. B. Bezerra, M. S. Cunha, L. F. F. Freitas and C. R. Muniz, Mod. Phys. Lett. A {\bf 32}, 1750005 (2017)], and the difference becomes crucial especially in the regions of near horizons of the spacetime.
\end{abstract}




\section{Introduction}

Motivated by the research papers devoted to the vacuum energy of a scalar massless field confined in a Casimir cavity moving in a circular equatorial orbit in the curved spacetime (see e.g. Refs.~\refcite{Karim} and~\refcite{Sorge} for the discussions in the Schwarzschild and Kerr space-time geometries, respectively) the authors of the recent paper~\cite{Bezerra} have calculated the Casimir energy of a massless scalar field in a cavity formed by nearby parallel plates orbiting a rotating gravitating object surrounded by quintessence, investigating the influence of the gravitational field on that energy, as measured by a comoving observer, with respect to whom the cavity is at rest, at zero temperature. The authors claim that the performed calculations include the effects due to the quintessence in addition to the dragging of inertial frames  for any radial coordinate and polar angle and according to their finding at the north pole the Casimir energy is not influenced by the quintessential matter.

This short note is devoted to the  recalculation of the Casimir energy of a massless scalar field in the spacetime of the Kerr black hole surrounded by quintessence and organized as the follows. Section~\ref{metric-QS} is devoted to the critical discussion of the failures of the spacetime metric~\cite{Ghosh} used and advantages of using of the spacetime metric~\cite{Toshmatov}.
In section~\ref{Casimir}, the Casimir energy in the Kerr spacetime with quintessence is recalculated. The last section summarizes our main results.

\section{On the spacetime metric of the Kerr black hole surrounded by quintessence}
\label{metric-QS}

In this section we provide detailed analysis of  the line element of the Kerr black hole surrounded by quintessential field~\cite{Ghosh} which has been used  in the paper~\cite{Bezerra} for studying the Casimir effect. It is well known that there was no rotating solution of the Einstein equations for the Kerr black hole surrounded by quintessence and only one remarkable solution of the Einstein equations for the spherically symmetric (Schwarzschild and Reissner-Nordstr\"{o}m) black holes in the quintessential field was obtained by Kiselev~\cite{Kiselev}. Afterwards, two different rotating quintessential black hole solutions (Kerr black hole in quintessence) have been obtained by converting so-called Kiselev solution (Schwarzschild black hole with quintessence) into rotational form by using the Newman-Janis algorithm (NJA) by Toshmatov et al.~\cite{Toshmatov} and Ghosh~\cite{Ghosh}. Now we provide comments on the latter one.

One of the ambiguous steps of the NJA is the complexification of the radial coordinate $r$~\cite{Newman-Janis,Bambi,Toshmatov1}, i.e.,
\begin{eqnarray}\label{complex}
r\rightarrow r+ia\cos\theta, \qquad u\rightarrow u-ia\cos\theta,
\end{eqnarray}
where $(u,r,\theta,\phi)$ are the Eddington-Finkelstein coordinates (EFCs). As a rule of the NJA, by using the above complex transformations the lapse function of the spherically symmetric spacetime metric takes new form, however, there is no unique way to complexification of the radial coordinate in the lapse function (for detail -- see~\cite{Azreg-Ainou}) and one can have as many different new forms of the lapse function as one wants.

One of the another drawbacks of the NJA is to bring the rotating spacetime metric from the EFCs to the Boyer-Lindquist coordinates (BLCs) by using the real coordinate transformations. In Ref.~\refcite{Ghosh} at the last step of the NJA in order to bring the rotating spacetime metric to the BLCs from the EFCs following coordinate transformations were used:
\begin{eqnarray}\label{01}
du=dt-Adr\ , \qquad d\phi=d\phi'-Bdr
\end{eqnarray}
where
\begin{eqnarray}\label{02}
A=\frac{r^2+a^2}{\Delta}\ , \qquad B=\frac{a}{\Delta}\ ,
\end{eqnarray}
with
\begin{eqnarray}\label{03}
\Delta=r^2+a^2-2Mr-\frac{\alpha}{\Sigma^{\frac{3\omega-1}{2}}}\ , \qquad \Sigma=r^2+a^2\cos^2\theta
\end{eqnarray}
The right hand sides of eqs.~(\ref{01}) are the total differentials (exact differentials) provided the functions $A$ and $B$ depend only on $r$. It is easy to check that, in this case, the conditions of integrability are satisfied, so one can integrate the two equations to obtain global coordinates $u(t,r)$ and $\phi(\phi,r)$. Unfortunately, one can see from expressions of the transformation functions~(\ref{02}) and~(\ref{03}) that this is not the case in the final expressions of $A$ and $B$ given in the right hand sides of eqs.~(\ref{01}), which generally depend on both ($r,\theta$). In this case the conditions of integrability are violated as following: If $A$ and $B$ depend on both ($r,\theta$) then $\partial A/\partial\theta\neq 0$ and $\partial B/\partial\theta\neq0$, so the conditions of integrability are no longer satisfied and it is not possible to integrate eqs.~(\ref{01}) with the transformation functions~(\ref{02}) and~(\ref{03}) to obtain global coordinates $u(t,r,\theta)$ and $\phi(t,r,\theta)$. Only in a few following special cases the conditions of integrability are satisfied: $\alpha=0$ and $a=0$ corresponds to the Schwarzschild solution, $\alpha=0$, which corresponds to the Kerr solution, and finally, $\alpha=-e^2$ and $\omega=1/3$ corresponds to the Kerr-Newman solution, which $A$ and $B$ depend only on $r$. However, in the case of black hole in the quintessential field the state parameter $\omega$ takes the value in the range $\omega\in(-1;-1/3)$. For these values of the state parameter $\omega$ transformation functions $A$ and $B$ remain to be dependent on both ($r,\theta$). These lead several problems such as violations of symmetry properties, impossibility of separations of variables in equations of motion, etc.

One more problem one comes across if the rotating solution of Ref.~\refcite{Ghosh} is used, we are sure, this solution does not satisfy the Einstein field equations $G_{\mu\nu}=8\pi T_{\mu\nu}$. Therefore, the claim ``solution" does not suit the spacetime metric (18) of Ref.~\refcite{Ghosh}.

Taking into account above comments, we have derived the another spacetime metric of the Kerr black hole in the quintessence that has very good agreement with above all comments (for detail -- see Ref.~\refcite{Toshmatov})
\begin{eqnarray}\label{metric}
ds^2=-\left(1-\frac{2\rho r}{\Sigma}\right)dt^2+\frac{\Sigma}{\Delta}dr^2&-&\frac{4a\rho r\sin^2\theta}{\Sigma}d\phi dt+\Sigma d\theta^2 \nonumber\\&&+\sin^2\theta\left(r^2+a^2+a^2\sin^2\theta\frac{2\rho r}{\Sigma}\right)d\phi^2\ ,
\end{eqnarray}
with
\begin{eqnarray}\label{notations}
\Sigma=r^2+a^2\cos^2\theta, \quad \Delta(r)=r^2-2\rho r+a^2, \quad 2\rho(r)=2M+\alpha r^{-3\omega} .
\end{eqnarray}
where $\alpha$ is the quintessential parameter representing intensity of the quintessence energy field related to the black hole and $\omega$ is quintessential equation of state (EoS) parameter.

\section{Casimir energy in the Kerr spacetime with
Quintessence}
\label{Casimir}

Recently, the Casimir effect has been studied in the Kerr spacetime surrounded by the quintessence by Bezerra et al.~\cite{Bezerra}. They have chosen the spacetime metric which was obtained in the paper~\cite{Ghosh} as the Kerr black hole in quintessence. However, we propose the spacetime~(\ref{metric}) is physically more favourite candidate for the Kerr black hole in the quintessential field. One can easily notice that these both two candidate spacetime metrics for the Kerr black hole with surrounded quintessential field are identical only on the equatorial plane, $\theta=\pi/2$.
\begin{figure}[t]
\begin{center}
\includegraphics[width=0.48\linewidth]{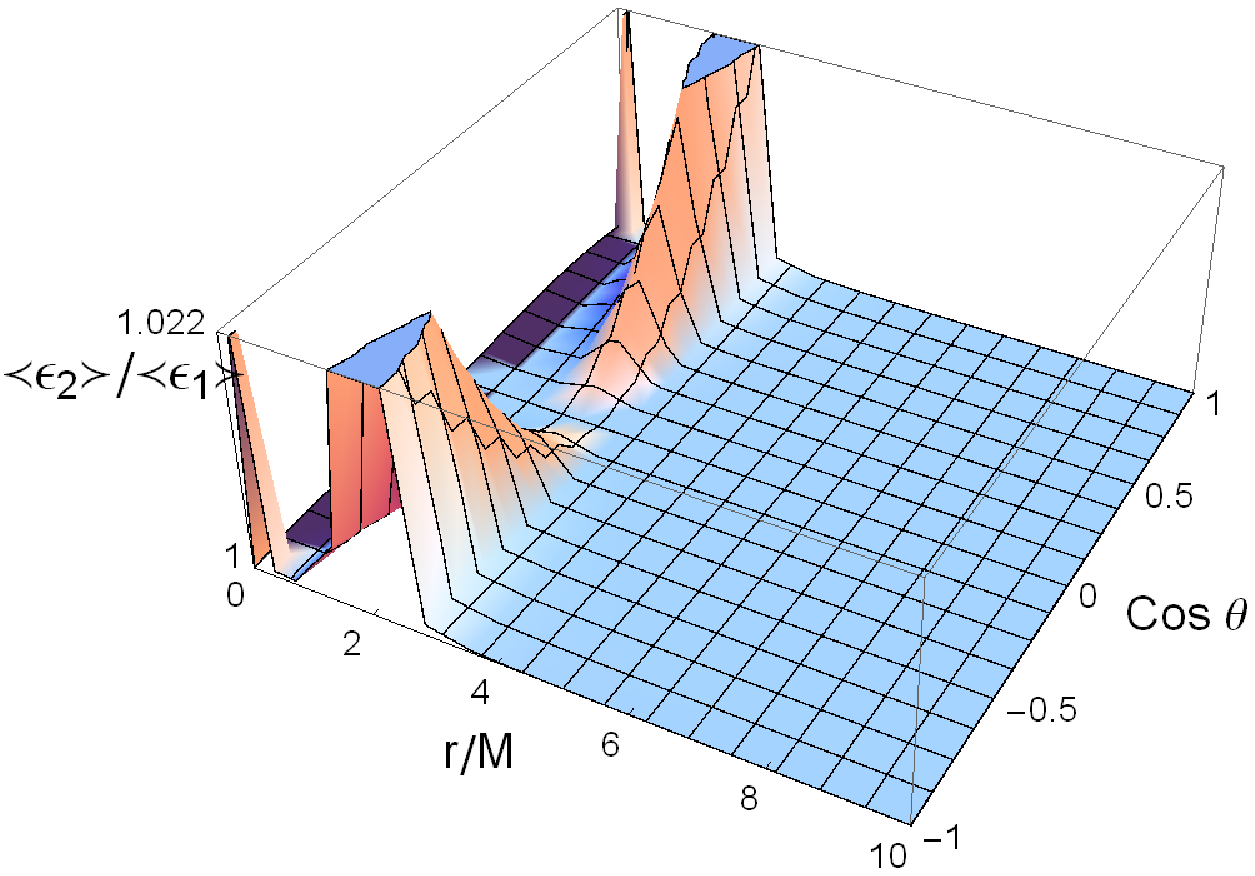}
\includegraphics[width=0.48\linewidth]{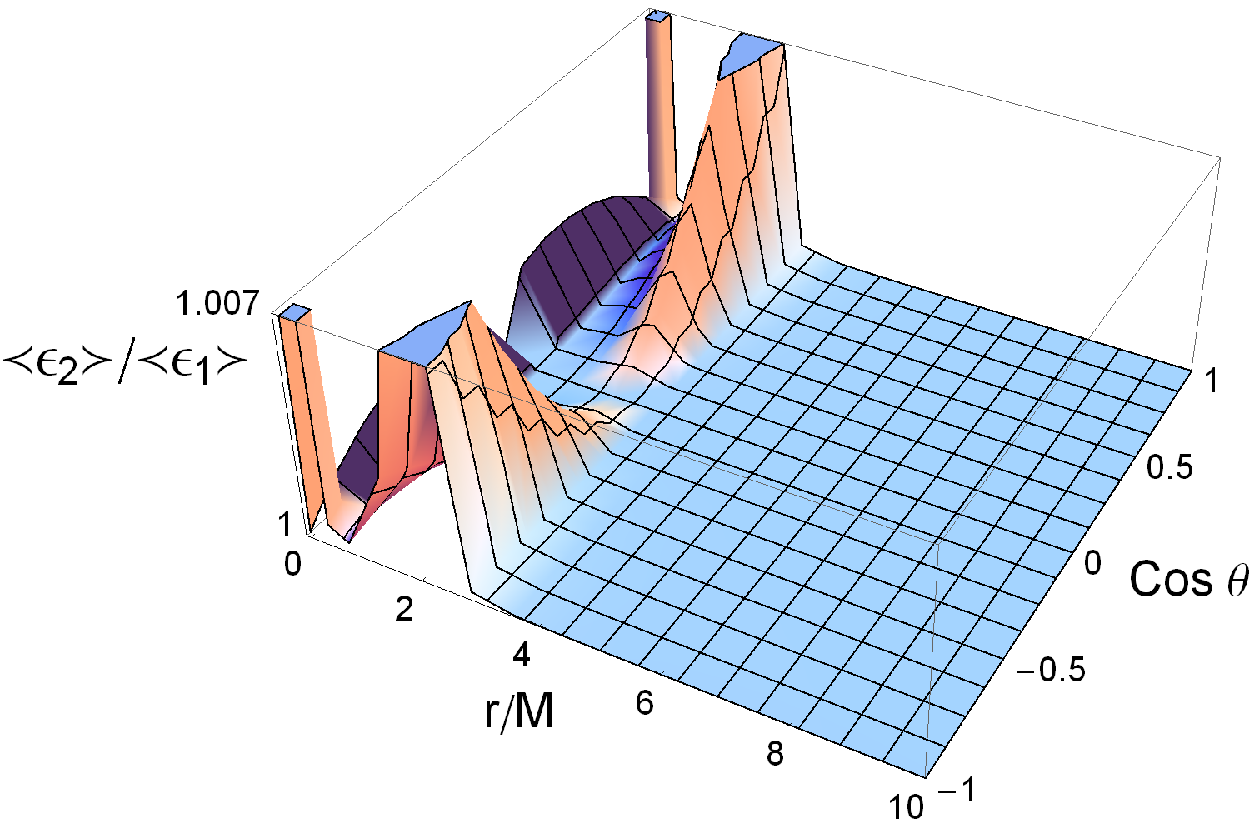}
\end{center}
\caption{\label{fig-ratio} Ratio of the Casimir energies in the quintessential Kerr black hole spacetimes Ref.~\cite{Toshmatov} (or spacetime metric (\ref{metric})), $<\epsilon_2>$, and Ref.~\cite{Ghosh}, $<\epsilon_1>$, for the values of the rotation and quintessence parameters $a/M=0.5$ and $\alpha M^2=0.01$, and the quintessential state parameter $\omega=-2/3$ (left) and $\omega=-2/5$ (right), respectively. In the left panel: inner and outer horizons are located at $r_-/M\approx0.13$ and $r_+/M\approx1.90$. In the right panel: $r_-/M\approx0.13$ and $r_+/M\approx1.89$.}
\end{figure}
In this section we aim to briefly compare the Casimir energy in the spacetime metric~(\ref{metric}) with the one of~\cite{Bezerra}. Here, we do not report full derivations of the expressions, since the all calculations have been presented in~\cite{Bezerra}. In Fig.~\ref{fig-ratio} the ratio of the near inner (Cauchy) and outer (event) horizons Casimir energies in these two spacetime metrics for the two typical values of the quintessence parameter are presented. One can see from Fig.~\ref{fig-ratio} that at the equatorial plane Casimir energies are the same in the both spacetimes. Note that the Casimir energy diverges at the coordinate singularities (horizons) of the spacetime.
\begin{figure}[t]
\begin{center}
\includegraphics[width=0.52\linewidth]{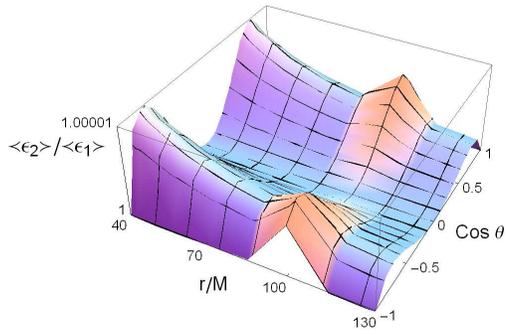}
\end{center}
\caption{\label{fig-ratio1} The same as Fig.~\ref{fig-ratio} but for the near quintessential horizon region. Where the quintessential horizon is located at $r_q/M\approx97.97$.}
\end{figure}
However, at the non-equatorial plane, near the inner and outer horizons the value of the Casimir energy in the spacetime~(\ref{metric}) starts to deviate (increase) from the one in the spacetime~\cite{Ghosh}. In Fig.~\ref{fig-ratio1} relative difference of the Casimir energies near the quintessential (cosmological) horizons of the both spacetimes is shown. As it is near the black hole horizons, near quintessential horizon also Casimir energy in the quintessential spacetime~(\ref{metric}) is bigger than one in Ref.~\refcite{Ghosh} ($<\epsilon_2>\geq<\epsilon_1>$) at the non-equatorial plane, $\theta\neq\pi/2$. At the spatial infinity Casimir energies in both spacetimes tend to one, since the both spacetimes are flat at large distances, close to the static radius.

\section{Conclusion}
\label{conclusion}

 The  main results of this short note can be summarized as the following.

We have performed the critical discussion of the failures of
the spacetime metric~\cite{Ghosh} and underlined the advantages of using of the spacetime metric~\cite{Toshmatov}.

 We have underlined the reasons for our study and explained the advantages of using the spacetime~\cite{Toshmatov} than the one~\cite{Ghosh} used in~\cite{Bezerra} and have shown that these spacetimes are identical only at the equatorial plane.

 The  recalculation of the Casimir energy of the massless scalar field in the more physically relevant spacetime of the Kerr black hole surrounded by quintessence derived in~\cite{Toshmatov} and its comparison with the results recently obtained in~\cite{Bezerra} in the spacetime~\cite{Ghosh}.

 We  have shown that in the more realistic spacetime which does not have the failures illustrated here the Casimir energy is significantly bigger than that derived by the authors of the recent paper~\cite{Bezerra} in the spacetime~\cite{Ghosh}, and the difference becomes crucial especially in the regions of near horizons of the spacetime, i.e. area which is very important for the astrophysical observations.

\section*{Acknowledgments}

B.T. and Z.S. acknowledge the Albert Einstein Center for Gravitation and Astrophysics supported by the Czech Science Foundation Grant No.~14-37086G and the Silesian University student grant SGS/14/2016. B.A. thanks the Goethe University, Frankfurt am Main, Germany, and the Faculty of Philosophy and Science, Silesian University in Opava, Czech Republic, for the warm hospitality. This research is supported in part by Grant No.~VA-FA-F-2-008 of the Uzbekistan Agency for Science and Technology,
by the Abdus Salam International Centre for Theoretical Physics through Grant No.~OEA-NT-01 and by the Volkswagen Stiftung, Grant No. 86 866.

\end{document}